\documentclass[runningheads]{llncs} 

\usepackage{mwe} 
\usepackage{bm}
\newcommand{\bb}[1]{\bm{\mathrm{#1}}}

\begin{document}

\title{3D FLAT: Feasible Learned Acquisition Trajectories for Accelerated MRI }

\author{Jonathan Alush-Aben 
\and Linor Ackerman-Schraier
\and Tomer Weiss 
\and Sanketh Vedula
\and Ortal Senouf
\and Alex Bronstein
}
\authorrunning{J. Alush-Aben et al.}
\institute{Technion - Israel Institute of Technology, Haifa 32000, Israel}

\maketitle

\begin{abstract}
Magnetic Resonance Imaging (MRI) has long been considered to be among the gold standards of today's diagnostic imaging. The most significant drawback of MRI is long acquisition times, prohibiting its use in standard practice for some applications. Compressed sensing (CS) proposes to subsample the $k$-space (the Fourier domain dual to the physical space of spatial coordinates) leading to significantly accelerated acquisition. However, the benefit of compressed sensing has not been fully exploited; most of the sampling densities obtained through CS do not produce a trajectory that obeys the stringent constraints of the MRI machine imposed in practice.
Inspired by recent success of deep learning-based approaches for image reconstruction and ideas from computational imaging on learning-based design of imaging systems, we introduce 3D FLAT, a novel protocol for data-driven design of 3D non-Cartesian accelerated trajectories in MRI. Our proposal leverages the entire 3D $k$-space to simultaneously learn a physically feasible acquisition trajectory with a reconstruction method. Experimental results, performed as a proof-of-concept, suggest that 3D FLAT achieves higher image quality for a given readout time compared to standard trajectories such as radial, stack-of-stars, or 2D learned trajectories (trajectories that evolve only in the 2D plane while fully sampling along the third dimension).  Furthermore, we demonstrate evidence supporting the significant benefit of performing MRI acquisitions using non-Cartesian 3D trajectories over 2D non-Cartesian trajectories acquired slice-wise.
\end{abstract}

\begin{keywords}
Magnetic Resonance Imaging, 3D MRI, fast image acquisition, image reconstruction, neural networks, deep learning, compressed sensing
\end{keywords}

\section{Introduction}
MRI is undoubtedly one of the most powerful tools in use for diagnostic medical imaging due to its noninvasive nature, high resolution, and lack of harmful radiation. It is, however, associated with high costs, driven by relatively expensive hardware and long acquisition times which limit its use in practice. 
Compressed sensing (CS) demonstrated that it is possible to faithfully reconstruct the latent images by observing a fraction of measurements \cite{candes2006robust}. In \cite{lustig2007sparse}, the authors demonstrated that it is theoretically possible to accelerate MRI acquisition by randomly sampling the $k$-space (the frequency domain where the MR images are acquired). However, many CS-based approaches have some practical challenges; it is difficult to construct a feasible trajectory from a given random sampling density or choose $k$-space frequencies under the constraints.


%
In addition, the reconstruction of a high-resolution MR image from undersampled measurements is an ill-posed inverse problem where the goal is to estimate the latent image $\bb{x}$ (fully-sampled $k$-space volume) from the observed measurements $\bb{y} = \mathcal{F}(\bb{x})+\bb{\eta}$, where $\mathcal{F}$ is the forward operator (MRI acquisition protocol) and $\bb{\eta}$ is the sampling noise. Some prior work approached this inverse problem by assuming priors (incorporated in a \textit{maximum a posteriori} setting) on the latent image such as low total variation or sparse representation in a redundant dictionary \cite{lustig2007sparse}. Recently, deep supervised learning based approaches have been in the forefront of the MRI reconstruction \cite{hammernik2018learning,sun2016deep,zbontar2018fastmri}, solving the above  inverse problem through implicitly learning the prior from a data set, and exhibiting significant improvement in the image quality over the explicit prior methods.  
Other studies, such as SPARKLING \cite{sparkling2019Lazarus}, have attempted to optimize directly over the feasible $k$-space trajectories, showing further sizable improvements. The idea of joint optimization of the forward (acquisition) and inverse (reconstruction) processes has been gaining interest in the MRI community for learning sampling patterns \cite{bahadir2019LOUPE}, Cartesian trajectories \cite{our2019fastmri,gozcu2018learning,zhang2019reducing} and feasible non-Cartesian 2D trajectories \cite{pilot2019weiss}.


%

%

%
 
 We distinguish recent works into four paradigms: (i) designing 2D Cartesian trajectories and sampling fully along the third dimension \cite{our2019fastmri,gozcu2018learning,zhang2019reducing}; (ii) designing 2D sampling densities and performing a full Cartesian sampling along the third dimension \cite{bahadir2019LOUPE}; (iii) designing {feasible} non-Cartesian 2D trajectories and acquiring slice-wise \cite{pilot2019weiss,sparkling2019Lazarus}; (iv) designing {feasible} non-Cartesian 3D trajectories, where the design space is unconstrained \cite{3dsparkling}. This work falls into the final paradigm.

 Cartesian sampling limits the degrees of freedom available in $k$-space data acquisition because it requires sampling fully along one of the dimensions. Acquiring non-Cartesian trajectories in $k$-space is challenging due to the need of adhering to physical constrains imposed by the machine, namely maximum slew rate of magnetic gradients and upper bounds on the peak currents. \cite{pilot2019weiss} developed a method for jointly training  2D image acquisition and reconstruction under these physical constraints, showing promising results and giving inspiration for this work. While there has been an attempt to optimize feasible 3D $k$-space trajectories in a follow up study on SPARKLING \cite{3dsparkling}, to the best of our knowledge, there has not been any research successful in exploiting the degrees of freedom available in 3D to design sampling trajectories by leveraging the strengths of data-driven reconstruction methods. This is the focus of the present study.

\paragraph{Contributions.} 
We propose 3D Feasible Learned Acquisition Trajectories (3D FLAT), a novel method for data-driven design of 3D non-Cartesian trajectories for MRI; simultaneously optimizing 3D $k$-space sampling trajectories with an image reconstruction method. We demonstrate that 3D FLAT achieves a significant improvement over standard 3D sampling trajectories -  radial and stack-of-stars \cite{radial-fov} - under a given time budget. We demonstrate the true merit of performing MRI acquisitions using non-Cartesian 3D trajectories over 2D non-Cartesian trajectories acquired slice-wise. Trajectories learned using 3D FLAT, in some cases, are able to accelerate acquisition by a factor of $2$ with no loss in image quality compared to the fixed trajectories using the same reconstruction method.

\section{The 3D FLAT Algorithm}
\begin{figure}
    \centering
    \includegraphics[width=\linewidth]{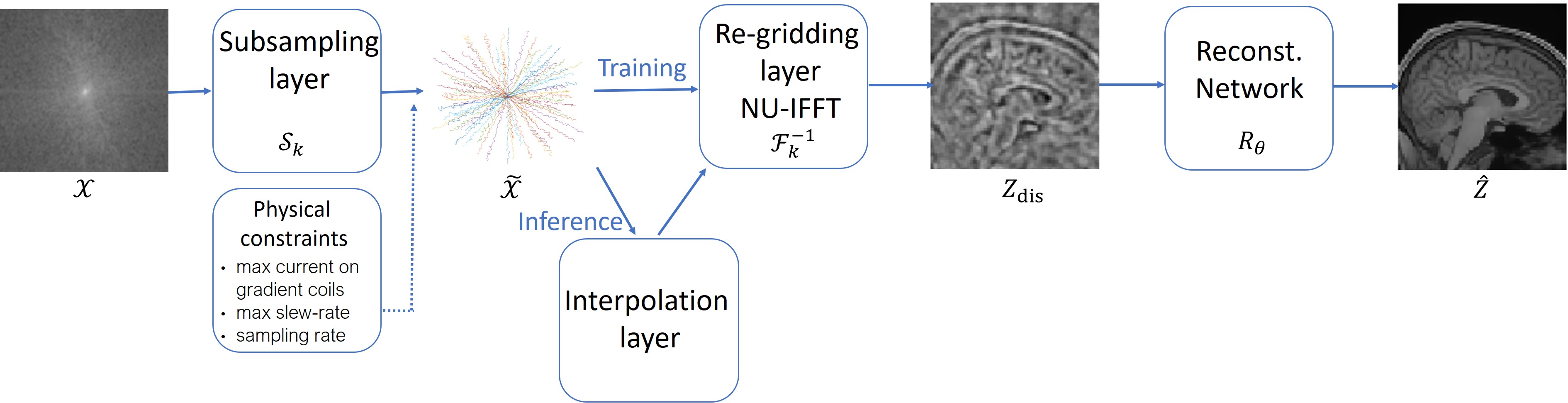}
    \caption{3D FLAT data flow pipeline. Notation is explained in the text.}
    \label{fig:model}
    \vspace{-0.5cm}
\end{figure}

Our algorithm can be seen as a pipeline combining the forward (acquisition) and the inverse (reconstruction) models ( Fig. \ref{fig:model}). The optimization of forward and inverse models is performed simultaneously while imposing physical constraints of the forward model through a penalty term in the loss function. The input to the forward model is the fully sampled $k$-space, followed by a sub-sampling layer, modeling the data acquisition along a $k$-space trajectory. The inverse model consists of a re-gridding layer, producing an image on a Cartesian grid in the spatial domain, and a convolutional neural network as a reconstruction model. 
%

\subsection{Forward model}
\paragraph{Sub-sampling layer.}
\label{subsec:subsampling}
The sub-sampling layer, $\mathcal{S_k}$, emulates MRI acquisition along the trajectory $\bb{k}$. The trajectory is a tensor $\bb{k}$ of size $N_{shots}\times m\times 3$; $N_{shots}$ is the number of RF (radio frequency) excitations, $m$ is the number of measurements per RF excitation, along three dimensions. The measurements form a complex vector of size $N_{shots}\times m$ emulated by bilinear interpolation $\tilde{\bb{x}} = \mathcal{S}_{\bb{k}} (\bb{X})$
on the full Cartesian grid $\bb{X} \in \bb{C}^{n \times n \times n}$ where the size of one dimension of the full Cartesian grid is denoted by $n$, W.L.O.G. assuming the full Cartesian grid is of size $n \times n \times n$.
A full Cartesian sampling consists of $n^2$ RF excitations, each for a line of the volume.
We refer to the ratio $\frac{n^2}{N_{shots}}$ as the \emph{acceleration factor} (AF). 

In order to obtain an efficient sampling distribution, 3D FLAT uses a coarsened trajectory containing $m'\ll m$ measurements per RF excitation, which is later interpolated to a trajectory of length $m$ with a cubic spline. This approach produces smooth results, well within the physical constraints of the MR machine and allows for efficient training. In addition to this practical advantage, we notice that updating the anchor points and then interpolating encourages more global changes to the trajectory when compared to updating all points. This is partly in spirit with observations made in \cite{boyer2016tsp} where linear interpolation is performed after reordering points using a traveling-salesman-problem solver.

\subsection{Inverse model}
\textit{Regridding layer.}
Conventionally, transforming regularly sampled MRI $k$-space measurements to the image domain requires the inverse fast Fourier transform (IFFT). However, the current case of non-Cartesian sampling trajectories calls for use of the non-uniform inverse FFT  (NuFFT) \cite{Dutt1993NUFFT}, $\hat{\mathcal{F}}^{-1}_{\bb{k}}$. The NuFFT performs regridding (resampling and interpolation) of the irregularly sampled points onto a grid followed by IFFT. The result is a (distorted) MR image,
$\bb{Z}_{dis} = \hat{\mathcal{F}}^{-1}_{\bb{k}} (\tilde{\bb{x}})$.


\textit{Reconstruction model.}
The reconstruction model extracts the latent image $\hat{\bb{Z}}$ from the distorted image $\bb{Z}_{dis}$; $\hat{\bb{Z}} = R_{\bb{\theta}}(\bb{Z}_{dis})$, $R$ represents the model and $\bb{\theta}$ its learnable parameters.
The reconstruction model passes the gradients back to the forward model in order to update the trajectory $\bb{k}$ so it will contribute most to the reconstruction quality. We emphasize that the principal focus of this work is not on the reconstruction model itself, and the proposed algorithm can be used with any differentiable model to improve the end-task performance.

\subsection{Loss function}
\label{subsec:loss}
The pipeline is trained by simultaneously learning the trajectory $\bb{k}$ and the parameters of the reconstruction model $\bb{\theta}$. To optimize the reconstruction performance while maintaining a feasible trajectory, we used a loss function composed of a data fidelity term and a constraint violation term, 
$L = L_{task} + L_{const}.$

\textit{Data fidelity.}
The $L_1$ norm is used to measure the discrepancy between the model output image $\hat{\bb{Z}}$ and the ground-truth image $\bb{Z}=\mathcal{F}^{-1}(\bb{X})$, derived from the fully sampled $k$-space,
$L_{task} = \| \hat{\bb{Z}} - \mathcal{F}^{-1}(\bb{X}) \|_1$.

\textit{Machine constraints.}
A feasible sampling trajectory must follow the physical hardware constraints of the MRI machine, specifically the peak-current (translated into the maximum value of imaging gradients $G_\mathrm{max}$), along with the maximum slew-rate $S_\mathrm{max}$ produced by the gradient coils. These requirements can be translated into geometric constraints on the first and second-order derivatives of each of the spatial coordinates of the trajectory:
$| \dot{k} | \approx \frac{| k_{i+1} - k_i |}{dt}  \le v_\mathrm{max} = \gamma\, G_\mathrm{max}$ and $| \ddot{k} | \approx \frac{| k_{i+1} - 2k_i + k_{i-1} | }{dt^2}  \le a_\mathrm{max} = \gamma \, S_\mathrm{max}$ ($\gamma$ is the gyromagnetic ratio).

The constraint violation term $L_{const}$ in the loss function applies to the trajectory $\bb{k}$ only and penalizes it for violation of the physical constraints. We chose the hinge functions of the form $\mathrm{max}(0, |\dot{k}|-v_\mathrm{max})$ and $\mathrm{max}(0, |\ddot{k}|-a_\mathrm{max})$ summed over the trajectory spatial coordinates and over all sample points. These penalties remain zero as long as the solution is feasible and grow linearly with the violation of each of the constraints. The relative importance of the velocity (peak current) and acceleration (slew rate) penalties is governed by the parameters $\lambda_v$ and $\lambda_a$, respectively. Note that in case of learning 3D trajectories, the constraints are enforced in 3 dimensions, corresponding to the respective gradient coils.

\textit{Optimization.} The training is carried out by solving the optimization problem 
\begin{equation}
\min_{\bb{k}, \bb{\theta}} \, \sum_{(\bb{X},\bb{Z})} L_{task}( R_{\bb{\theta}}(\hat{\mathcal{F}}^{-1}_{\bb{k}}(\mathcal{S}_{\bb{k}} (\bb{X}) ) ) , \bb{Z}) + L_{const}(\bb{k}),
\label{eq:min}
\end{equation}
where the loss is summed over a training set comprising the pairs of fully sampled data $\bb{X}$ and the corresponding groundtruth output $\bb{Z}$.

\section {Experimental evaluation}
Our code is available at \url{https://github.com/3d-flat/3dflat}.
\subsection{Dataset}
T1-weighted images taken from the human connectome project (HCP) \cite{HCP@2012} were used. We down-sampled the HCP's 1065 brain MRI volumes to $80\times80\times80$, from the original $145\times174\times145$, keeping an isotropic spatial resolution, and using a 90/10 split for training/validation.

\subsection{Training settings}

The network was trained using the Adam \cite{adam2014} optimizer. Learning rate was set to $0.001$ for the reconstruction model, and $0.005$ for the sub-sampling layer. For the differentiable regridding layer (Nu-IFFT) \cite{Dutt1993NUFFT}, we made use of an initial 2D implementation available from the authors of \cite{pilot2019weiss}\footnote{\url{https://github.com/tomer196/PILOT}}. For the reconstruction model, we used a 3D U-Net architecture \cite{unet3d}, based on the publicly-available implementation\footnote{\url{https://github.com/wolny/pytorch-3dunet}}. U-Net is widely-used in medical imaging tasks in general, and in MRI reconstruction \cite{zbontar2018fastmri} and segmentation \cite{MRIsegment} in particular.
We emphasize that the scope of this work is not directed toward building the best reconstruction method, but rather demonstrating the benefit of simultaneous optimization of the acquisition-reconstruction pipeline; any differentiable reconstruction method is suitable.
The physical constraints we enforced are: $G_\mathrm{max}=$ 40mT/m for the peak gradient, $S_\mathrm{max}=$ 200T/m/s for the maximum slew-rate, and $dt = 10\mu$sec for the sampling time.

\subsection{Reference trajectories}
Standard trajectories used in 3D multi-shot MR imaging are radial lines and stacks of stars (SOS) \cite{glover1992radial}. SOS is a 2D radial trajectory in the $xy$ plane multiplexed over the $z$ dimension. In the experiment \emph{SOS 2D}, a trajectory was initialized with SOS and learned in the $xy$ plane only.
We claim 3D-FLAT can optimize any heuristically hand-crafted trajectory, but limit our choice to these radial initializations. Other trajectories such as Wave-CAIPI as proposed in \cite{waveCAIPI}, could be used as well.
All slices were acquired with the same trajectory as in the multi-shot PILOT experiment suggested in \cite{pilot2019weiss}. In the experiment \emph{SOS 3D}, a trajectory initialized with SOS was allowed to train in 3D, exploring all degrees of freedom available.The 3D radial trajectories were constructed as described in \cite{radial-fov}, using the MATLAB implementation\footnote{\url{https://github.com/LarsonLab/Radial-Field-of-Views}}. The 2D radial trajectories were evenly distributed around the center.
In our simulation, sampling $m=3000$ data-points over per shot of $N_{shots}$ did not add any new information, unlike a real sampling scenario. For ease of computation, $m'=100$ points were sampled.

\begin{figure}[!htb]
   \centering
   \addtolength{\tabcolsep}{5.5pt}
\begin{tabular}{c c c}
\raisebox{-0.5\height}{\includegraphics[height=3cm, width=3cm, angle=90]
{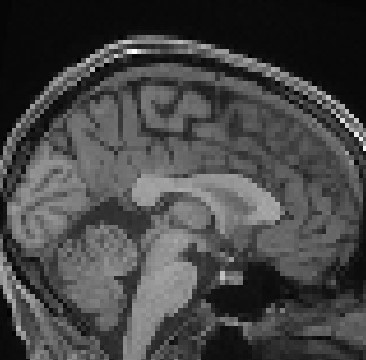}}&

\raisebox{-0.5\height}{\includegraphics[height=3cm, width=3cm, angle=90]
{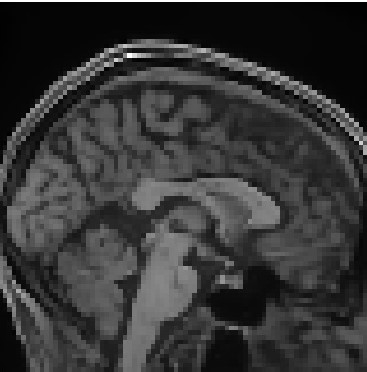}}
&
\raisebox{-0.5\height}{\includegraphics[height=3cm, width=3cm, angle=90]
{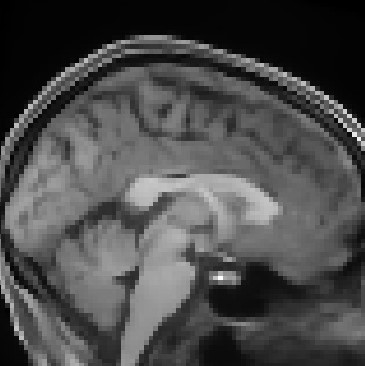}}
\\
\raisebox{-0.5\height}
{\includegraphics[width=0.3\textwidth, height=0.32\textwidth]{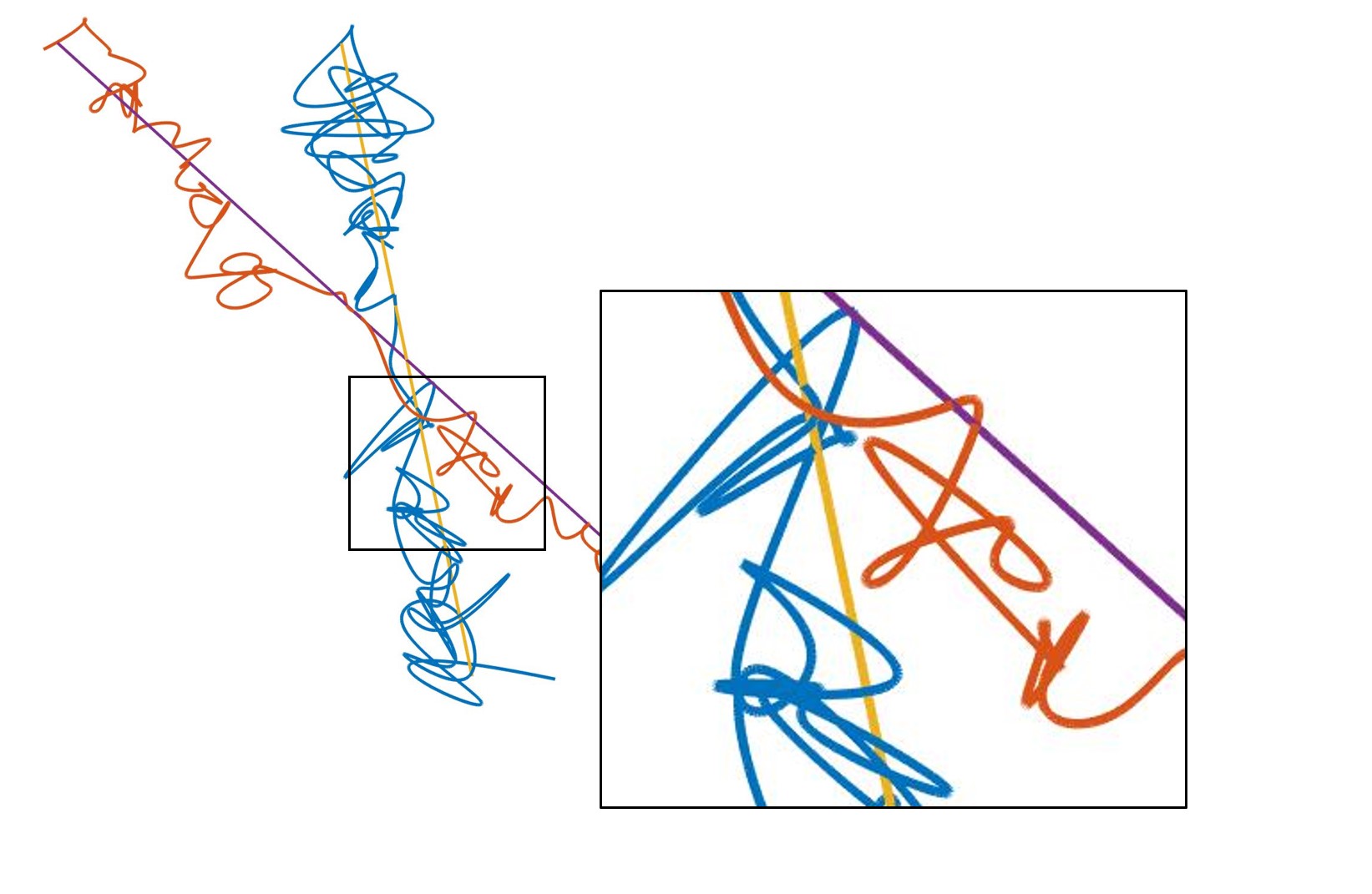}} & 

\raisebox{-0.5\height}{\includegraphics[width=0.3\textwidth, height=0.32\textwidth]{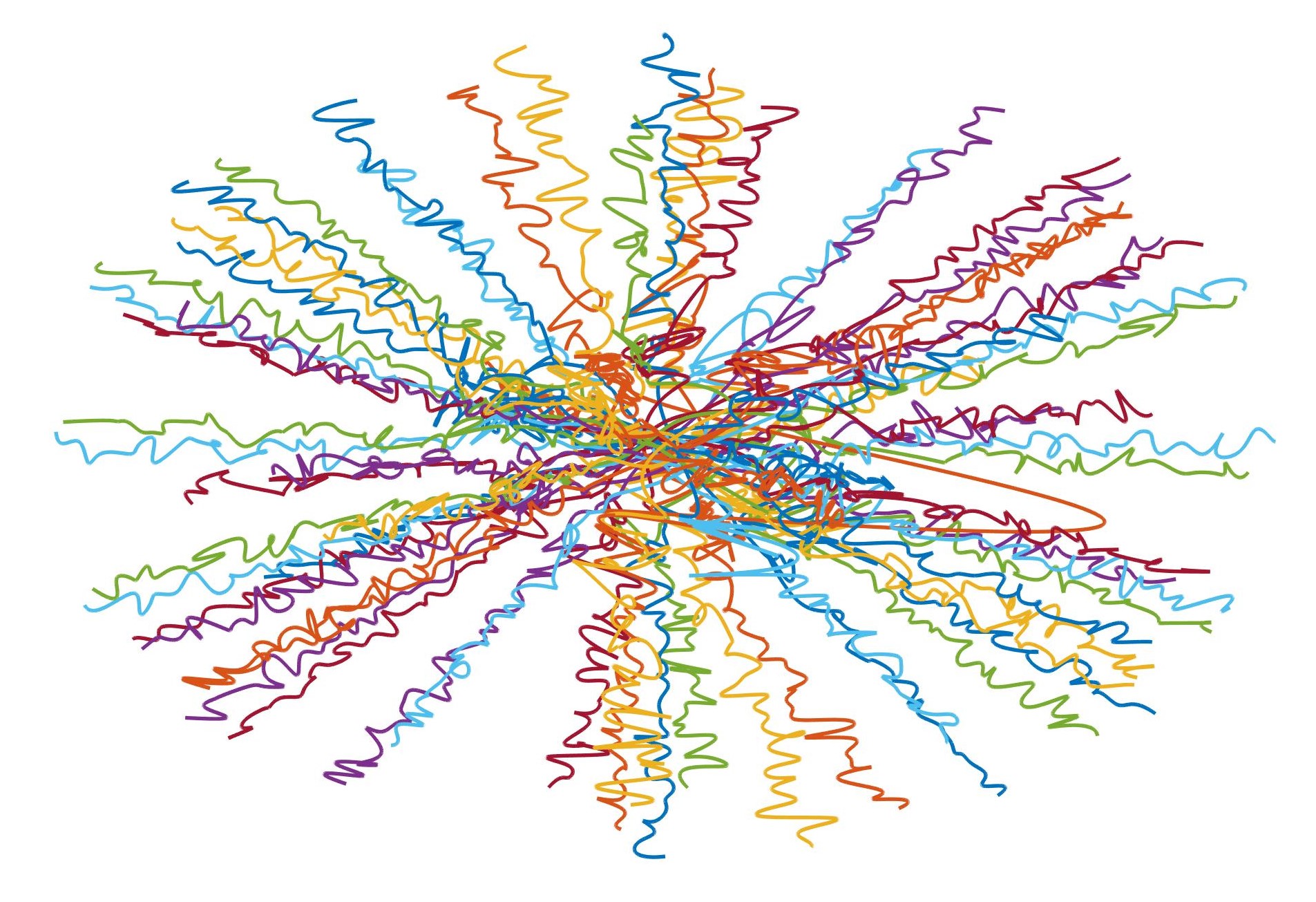}}& 
\raisebox{-0.5\height}
{\includegraphics[width=0.3\textwidth, height=0.32\textwidth]{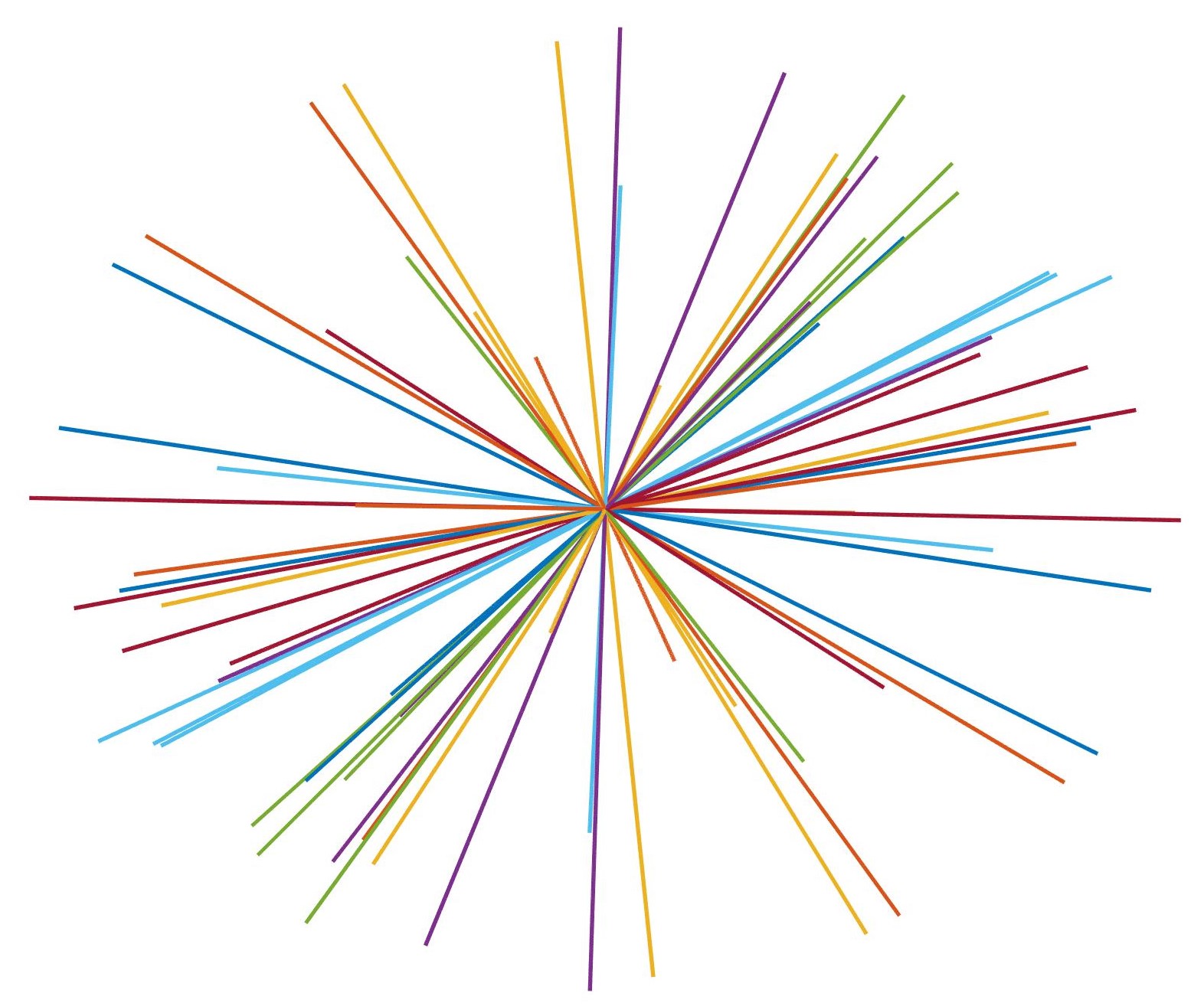}}
\\

\end{tabular}
\vspace{-0.2cm}
\caption{First row: reconstruction results using different sampling methods (left-to-right): groundtruth image using full $k$-space; 3D FLAT with radial initialization; and fixed radial trajectory. Second row: depiction of the trajectories (left-to-right): 2 shots of 3D FLAT learned trajectory and its initialization overlaid; 3D FLAT learned trajectory; and radial trajectory used for initalizing for 3D FLAT. Note that the second row presents 3D trajectories but visualized in 2D.}
\label{fig:withtraj_mainPaper}
\vspace{-0.9cm}
\end{figure}

\subsection{Results and discussion}

For quantative evaluation, we use the peak signal-to-noise ratio (PSNR) and structural-similarity (SSIM) \cite{ssim} measures. All trajectories used in our experiments are feasible, satisfying mentioned machine constraints.
\begin{figure}
\includegraphics[width=1\textwidth,trim={1cm 0 0.5cm 0},clip]{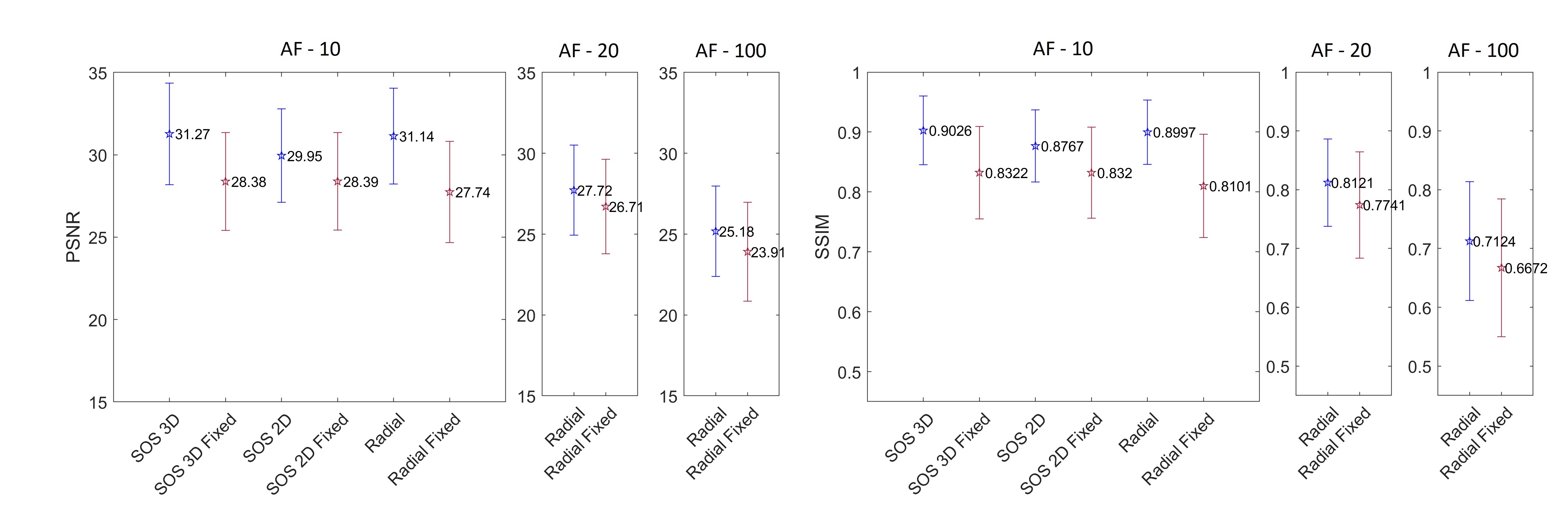}
\vspace{-1cm}
\caption{Quantitative results (PSNR \& SSIM) of 3D FLAT for different acceleration factors and initializations. 3D FLAT outperforms fixed trajectories in all acceleration factors and initialization. Error bars report best, worst and average values seen.
}
\label{psnr-cmp}
\vspace{-0.7cm}
\end{figure}
We compared our algorithms to training a reconstruction model using measurements obtained using fixed handcrafted trajectories, ``fixed trajectories". Quantitative results (Fig. \ref{psnr-cmp}) show that every 3D FLAT trajectory (the learned SOS 2D/3D and radial) outperforms fixed trajectories in every acceleration factor (AF). We notice an improvement of $1.01-3.4$ dB in PSNR and $0.0452-0.0896$ in SSIM on full 3D trajectories. The standard deviation over the SSIM metric is noticeably smaller in some cases of 3D FLAT trajectories. Notice in Fig. \ref{psnr-cmp}, achieved mean PSNR of the learned radial trajectory with AF of 20 performs similarly to that of a fixed radial trajectory with AF of 10, yielding a speedup factor of $2$ with no loss in image quality. Furthermore, results suggest that learning the trajectory in all three dimensions (learned SOS 3D and learned radial) leads to a more significant improvement ($1.32$ dB in PSNR) over the same initialization allowed to learn in two dimensions. This corroborates the natural assumption that an additional degree of freedom (3D) used to design trajectories can improve the end image quality. This improves the result of 3D SPARKLING \cite{3dsparkling} where, to the best of our understanding, the authors reached a conclusion that 2D SPARKLING trajectories acquired slice-wise outperformed the 3D ones. The possible limitations of 3D SPARKLING which 3D FLAT alleviates are twofold: Firstly, 3D SPARKLING enforces constraints on the search space of feasible trajectories by replicating the same learned trajectory in multiple regions of the $k$-space whereas in 3D FLAT the search space is unconstrained. Secondly, 3D SPARKLING requires an estimate of the desired sampling density prior to optimization whereas 3D FLAT enables task-driven learning of optimal sampling density jointly with the feasible trajectories (Fig. \ref{fig:sampling_density}).

Qualitative results in Figs. \ref{fig:3dflat_teaser}, \ref{saggitalpics} \& \ref{coronalpics} present visual depiction of sample slices from multiple views obtained using different acquisition trajectories (learned and fixed) at different AFs. The visual results suggest that learned trajectories contain more details and are of superior quality than the fixed counterparts. That said, performing experiments on real machines is necessary for the next steps of the research, but out of the scope of this proof-of-concept.

We tested our algorithm across three AFs, 10, 20, and 100, where they demonstrated invariable success in improving the reconstruction accuracy. We notice the radial trajectory with AF $20$ performs as well as the fixed radial trajectory of AF $10$.
To demonstrate the robustness of 3D FLAT to different reconstruction methods, we performed reconstructions using off-the-shelf TV-regularized compressed-sensing inverse problem solvers \cite{espiritbart}. The results are presented in Table 1 in the Appendix comparing 3D FLAT with fixed counterparts. In all cases, 3D FLAT outperformed fixed trajectories in terms of PSNR.

\textit{Learned trajectories and sampling densities.} Visualizations of learned trajectories with different initializations are presented in Figs. \ref{fig:withtraj_mainPaper}, \ref{fig:vis_traj}, and \ref{fig:singleshot}, Different shots are coded with different colors. Fig. \ref{fig:withtraj_mainPaper} (bottom row) depicts the learned and fixed radial trajectories, Fig. \ref{fig:vis_traj} shows the learned SOS 2D and 3D trajectories plotted with the fixed SOS trajectory. Note that the learned SOS 2D and 3D trajectories might look similar due to visualization limitations. A close up of one shot is presented in Fig. \ref{fig:singleshot}; it is interesting to notice the increased sampling density in high-curvature regions of the trajectory. The reason could be due to the enforced constraints on the slew-rate which do not allow sharp turns in the trajectory, resulting in increased sampling in these regions. Furthermore, a visualization of sampling density of learned and fixed trajectories is presented in Fig. \ref{fig:sampling_density}. We see the learned radial and SOS 3D have improved sampling density at the center of the $k$-space. This coincides with the intuition suggesting that the center containing more low-frequency information is more important for reconstruction. This is a possible reason for 3D FLAT outperforming the fixed trajectories while using TV-regularized direct reconstruction. 
We also notice a dependence of the learned trajectories on initialization, this problem was also encountered by \cite{sparkling2019Lazarus} and \cite{pilot2019weiss}, this could be due to local minima near them.



\section{Conclusion}

We demonstrated, as a proof-of-concept, that learning-based design of feasible non-Cartesian 3D trajectories in MR imaging leads to better image reconstruction when compared to the off-the-shelf trajectories. To the best of our knowledge, this is the first attempt of data-driven design of feasible 3D trajectories in MRI. We further demonstrate the benefit of acquiring 3D non-Cartesian trajectories over their 2D counterparts acquired slice-wise. Our experiments suggest that the learned trajectories fall significantly below the enforced machine constraints. We believe that such trajectories can be deployed in low magnetic field portable MRI scanners (i.e. Hyperfine\footnote{\url{https://www.hyperfine.io/}}) to achieve better reconstruction accuracy vs. acquisition speed. We plan to try this in future work. We defer the following aspects to future work:
Firstly, in this work, we limited our attention to a relatively small resolution of the $k$-space. The reason for this is due to the computational complexity of our reconstruction method which can be alleviated by using 3D CNNs with lower complexity. 
Secondly, this work did not take into account signal decay within a single acquisition. This noise can be modelled and taken into account during training. However, since each shot is relatively small ($30ms$), we believe there would not be a noticeable effect on the final reconstruction.
Thirdly, the trajectories achieved are sub-optimal, as they are highly dependant on initialization. This could be improved by optimizing in a two step process: optimizing for the optimal sampling density and then designing a trajectory while enforcing machine constraints, this was proposed in \cite{pilot2019weiss} in a 2D single-shot scenario (PILOT-TSP).
Lastly, this work is a proof-of-concept that has been validated through simulations and has yet to be validated on real MRI machines.





\bibliographystyle{splncs04}
\bibliography{3dflat}

\newpage
\appendix

\section{Supplementary material}
\label{appendix}

\begin{figure}[!b]
   \centering
\begin{tabular}{c c c c}
& Radial  & SOS 3D & SOS 2D \\
\rotatebox[origin=c]{90}{Fixed} & \raisebox{-0.5\height}{
\includegraphics[width=0.31\textwidth, height=0.29\textwidth]{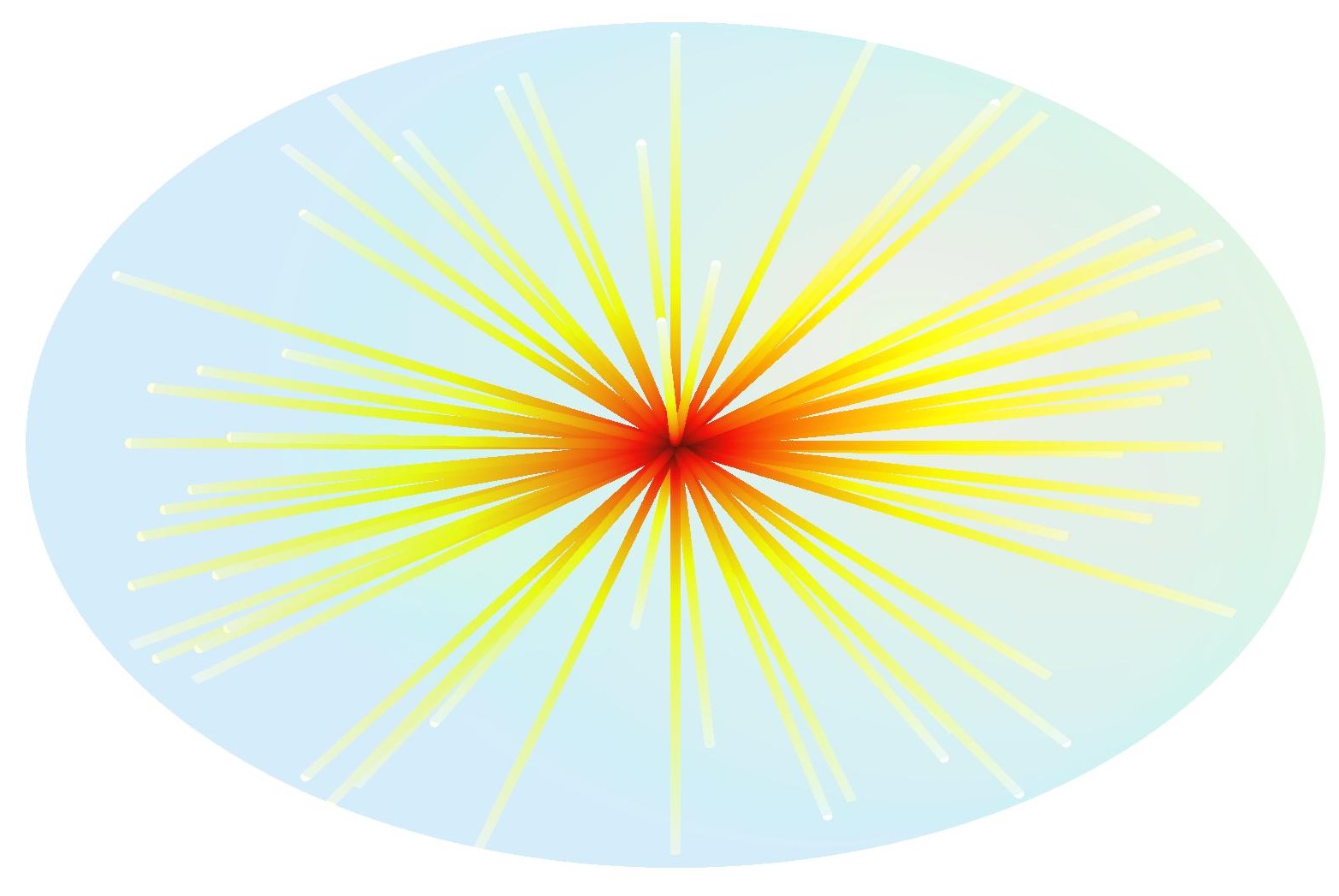}} &
 \raisebox{-0.5\height}{\includegraphics[width=0.3\textwidth, height=0.31\textwidth]{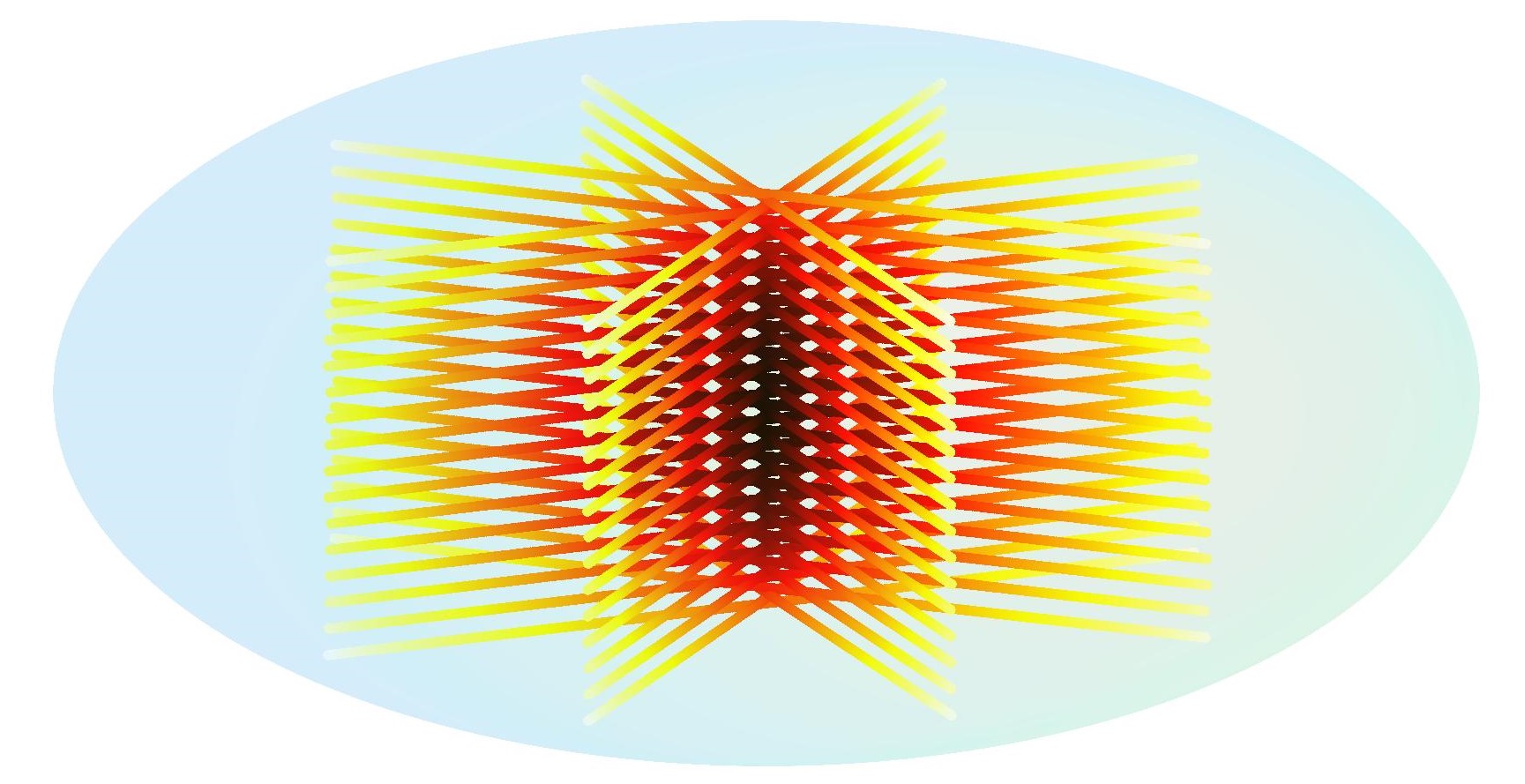}} &
 \raisebox{-0.5\height}{\includegraphics[width=0.3\textwidth, height=0.31\textwidth]{figs/densit_map/sosF-bigsphere.jpg}} \\
 \rotatebox[origin=c]{90}{Learned} &
\raisebox{-0.5\height}{\includegraphics[width=0.3\textwidth, height=0.29\textwidth]{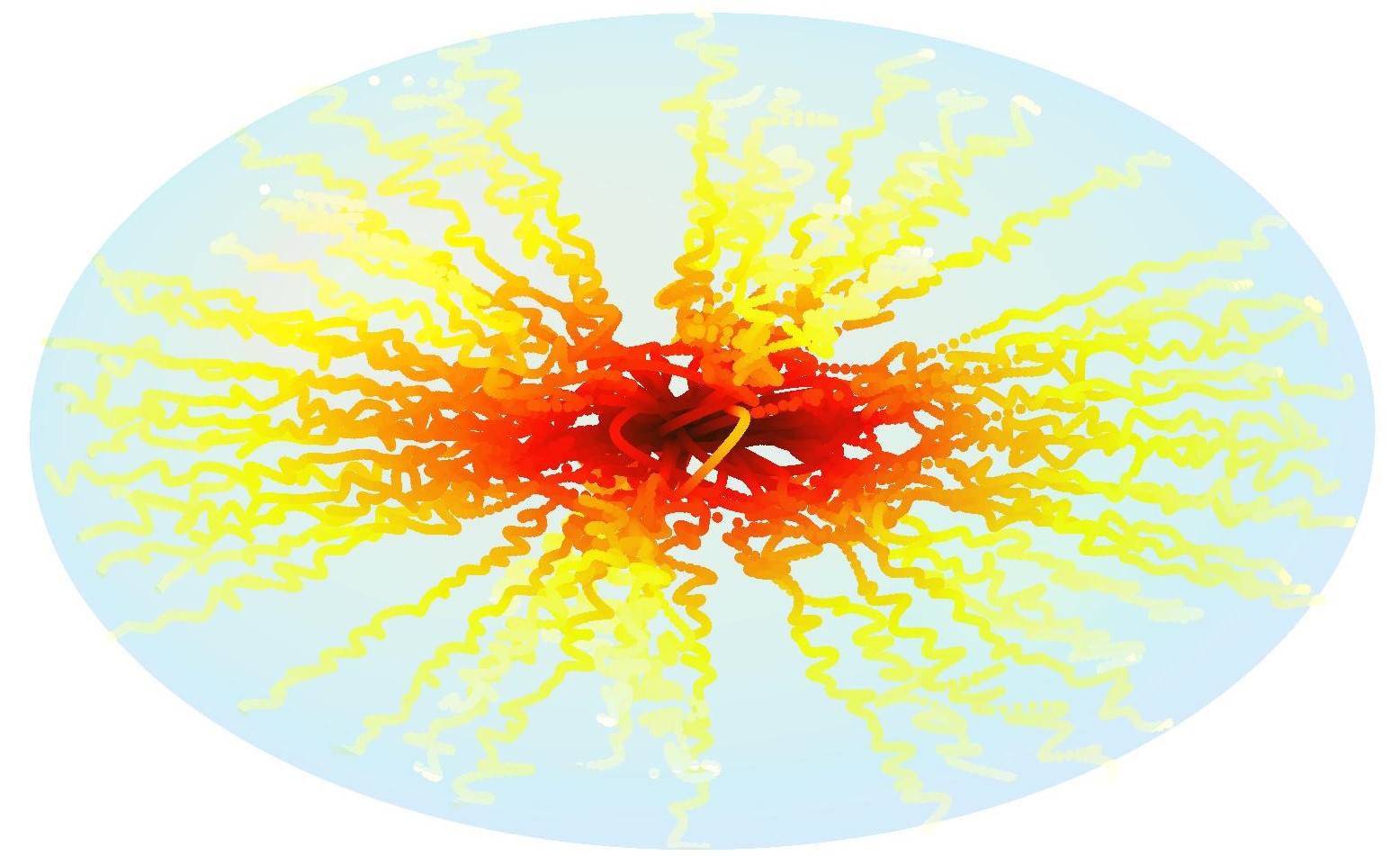}}&
\raisebox{-0.5\height}{\includegraphics[width=0.3\textwidth, height=0.32\textwidth]{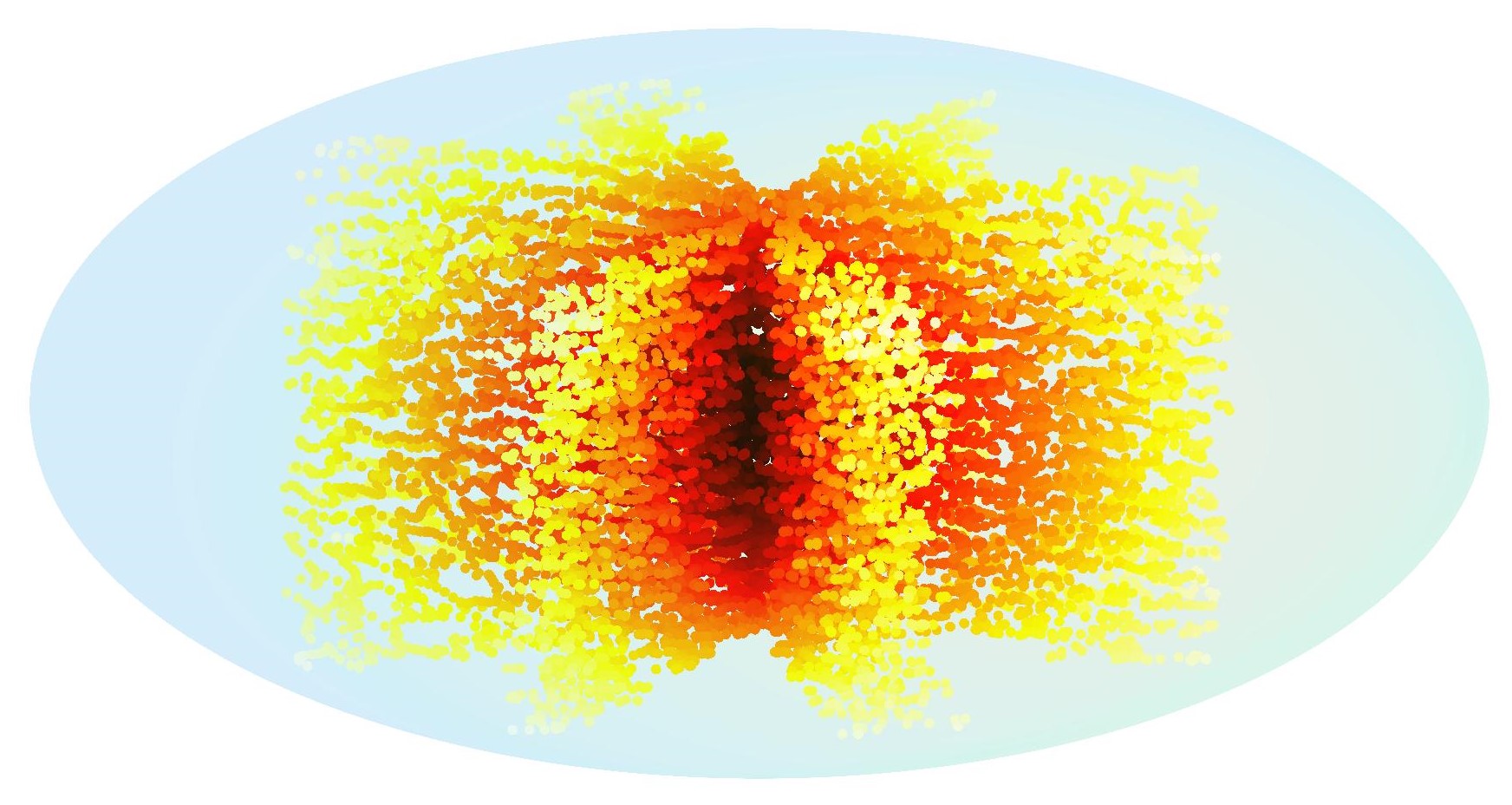}}&
\raisebox{-0.5\height}{\includegraphics[width=0.3\textwidth, height=0.32\textwidth]{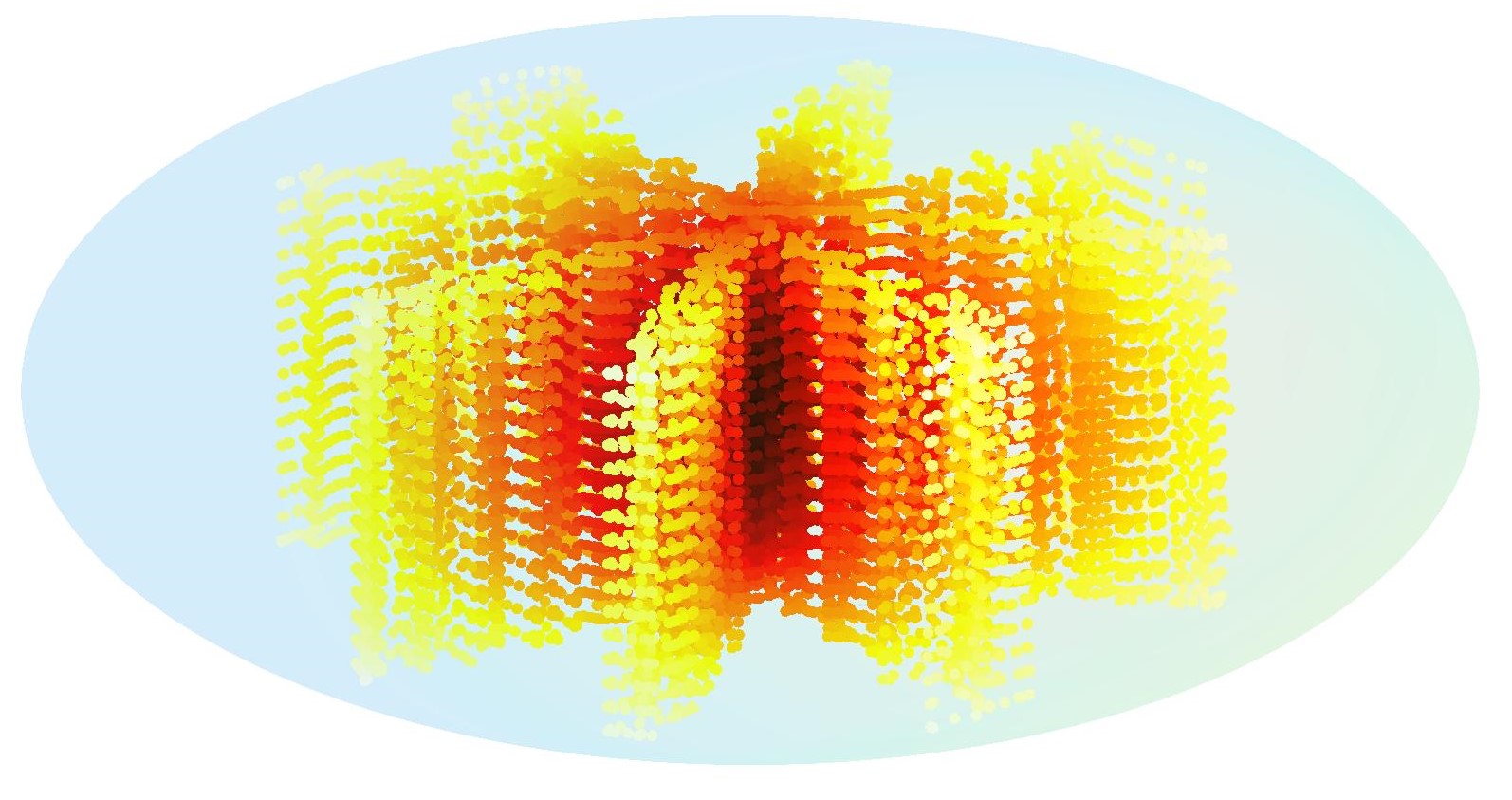}}\\
\end{tabular}   

\caption{The sampling densities of the fixed trajectories and 3D FLAT are visualized above. Notice the change in density between the fixed initialization and the learned. Note that for visualization purposes only a fraction of the points are shown. Best viewed in color.}
\label{fig:sampling_density}   
\end{figure}

\begin{figure}[!htb]
   \centering

\begin{tabular}{c c c}

SOS & Learned SOS 2D & Learned SOS 3D \\
\includegraphics[, trim={4cm 0 4cm 0},clip,width=0.30\textwidth, height=0.32\textwidth]{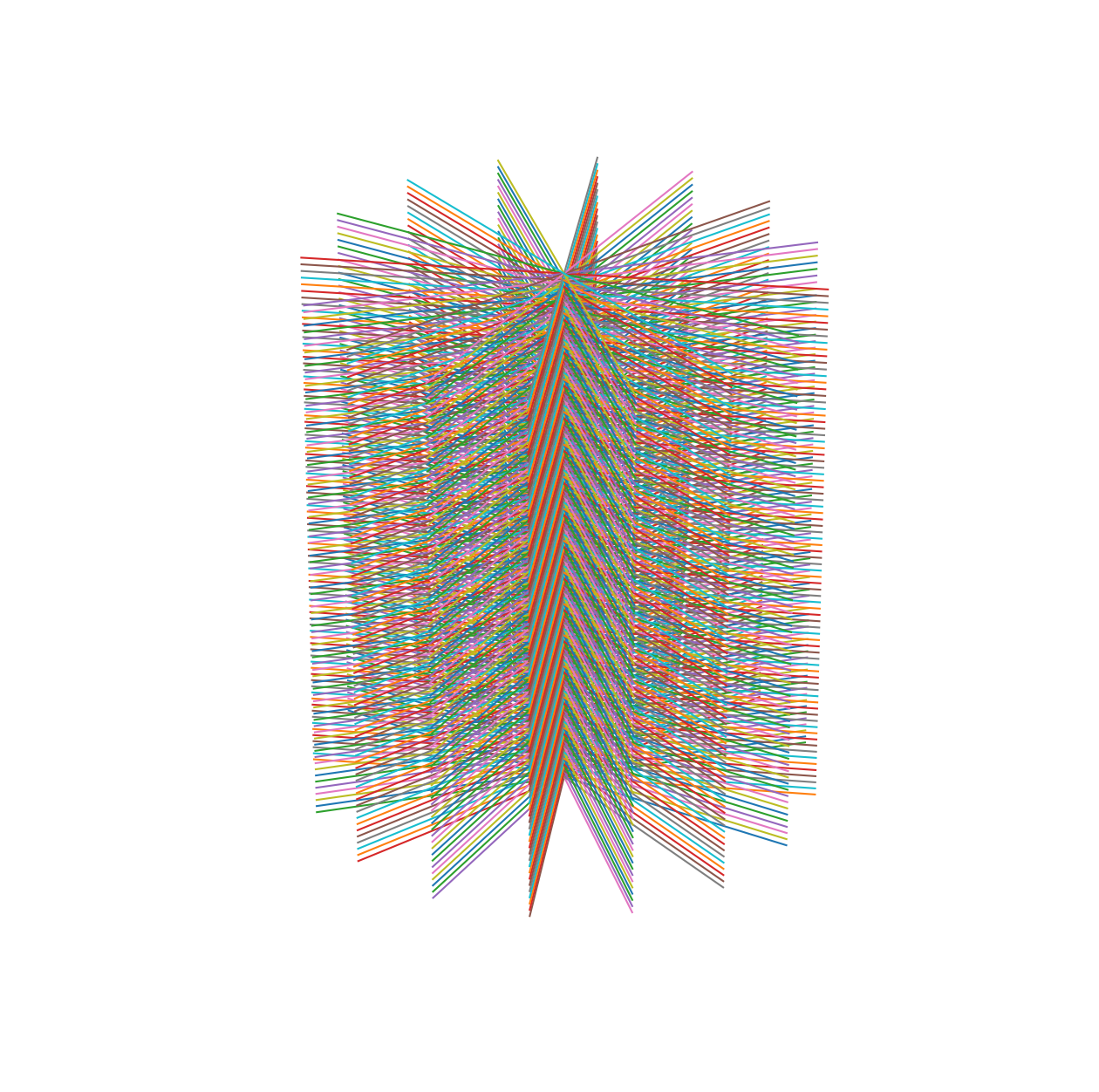}&
\includegraphics[width=0.30\textwidth, height=0.32\textwidth]{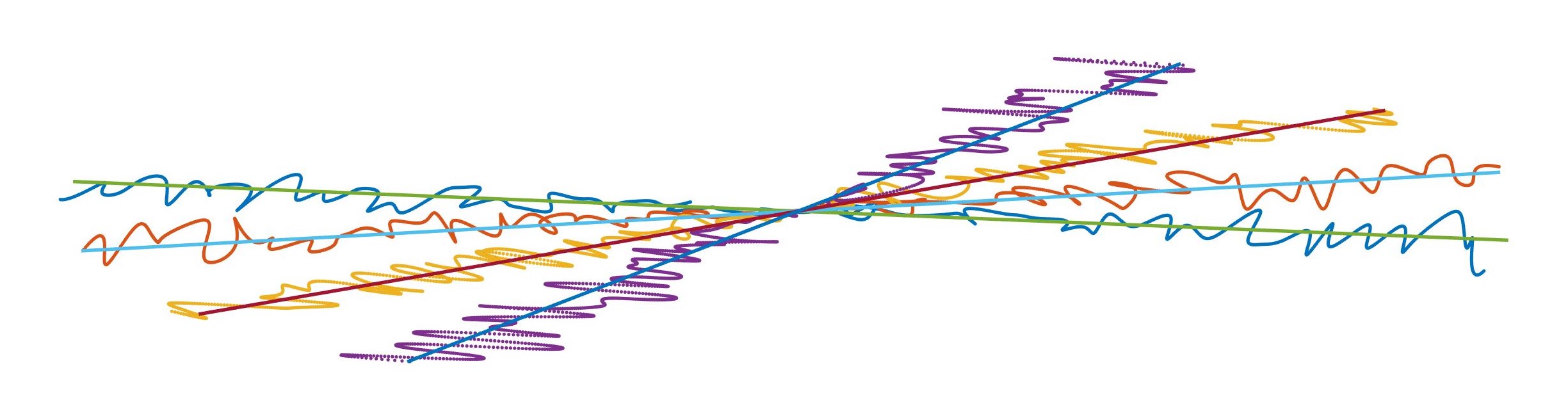}&
\includegraphics[width=0.30\textwidth, height=0.32\textwidth]{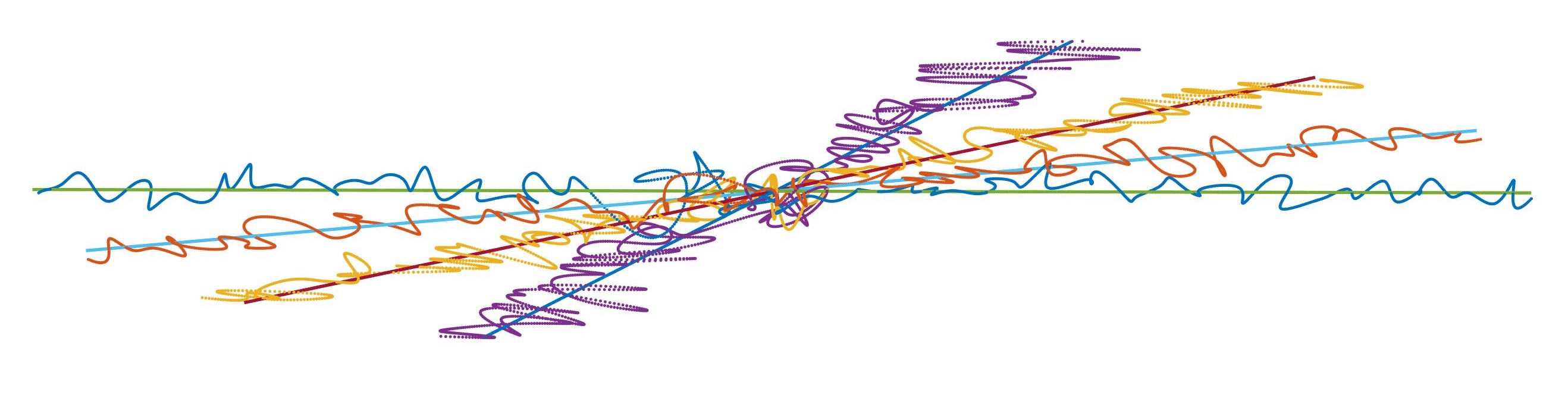}
\\
\end{tabular}
\caption{Compared above are three trajectories initialized using stack-of-stars. From left to right are: fixed trajectory; 4 shots of learned SOS 2D trajectories; 4 corresponding shots of learned SOS 3D trajectories. Note that the right most trajectory is actually 3D and that all the trajectories obey the MR physical constraints.}
\label{fig:vis_traj}
\end{figure}

\begin{table}[!htb]
 \centering
  {tab:cs}%
  {\begin{tabular}{c c c c}
  \bfseries Trajectory &\bfseries Acceleration Factor& \bfseries Fixed & \bfseries Learned\\
  Radial & 10  &17.19 &17.99\\
  Radial & 20 &16.98 &17.24\\
  Radial &100 & 16.75 & 17.5\\
  SOS 3D &10 & 16.14 &16.98
  \end{tabular}}
  {\caption{Comparison of 3D FLAT and fixed trajectories with TV-regularized image reconstruction using off-the-shelf CS inverse problem solvers (BART, \cite{espiritbart}). Learned trajectories outperform the fixed counterparts across all acceleration factors and initializations.}}%
  \label{TVResults}
\end{table}

\begin{figure}[!b]
\addtolength{\tabcolsep}{5.5pt}
   \centering
\begin{tabular}{c c c c}
& Sagittal  & Coronal & Axial \\
\rotatebox[origin=c]{90}{Groundtruth} & 
\raisebox{-0.5\height}{\includegraphics[width=3cm, height=3cm, angle=90]{figs/brain/sagittal_img_113619_orig.jpg}}&
\raisebox{-0.5\height}{\includegraphics[width=3cm, height=3cm, angle=90]{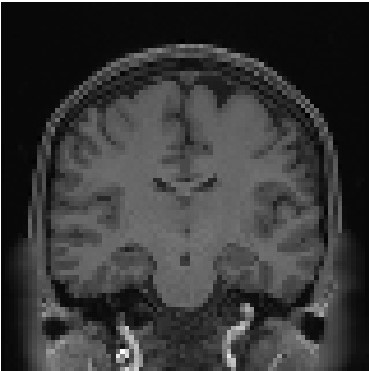}}&
\raisebox{-0.5\height}{\includegraphics[width=3cm, height=3cm, angle=90]{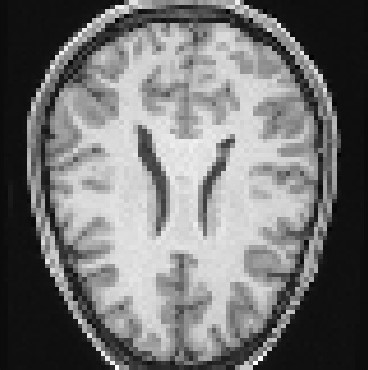}} 
\\ \\
\rotatebox[origin=c]{90}{3D FLAT} & 
 \raisebox{-0.5\height}{\includegraphics[width=3cm, height=3cm, angle=90]{figs/brain/sagittal_img_113619_learned_reconstructed.jpg}}&
\raisebox{-0.5\height}
{\includegraphics[width=3cm, height=3cm, angle=90]{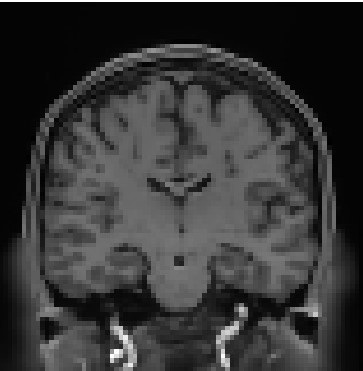}}&
\raisebox{-0.5\height}
{\includegraphics[width=3cm, height=3cm, angle=90]{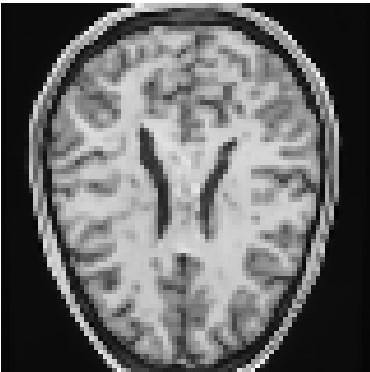}}\\

\end{tabular}   
\caption{The space is processed in 3D, any plane can be depicted easily. Shown above are three planes of the same volume. The first row is the ground truth image, processed with the full k-space. The second was created with 3D FLAT initialized with a radial trajectory at an acceleration factor of 10.}
\label{fig:3dflat_teaser}
\end{figure}

\begin{figure}[!htb]
   \centering
\includegraphics[width=0.9\textwidth]{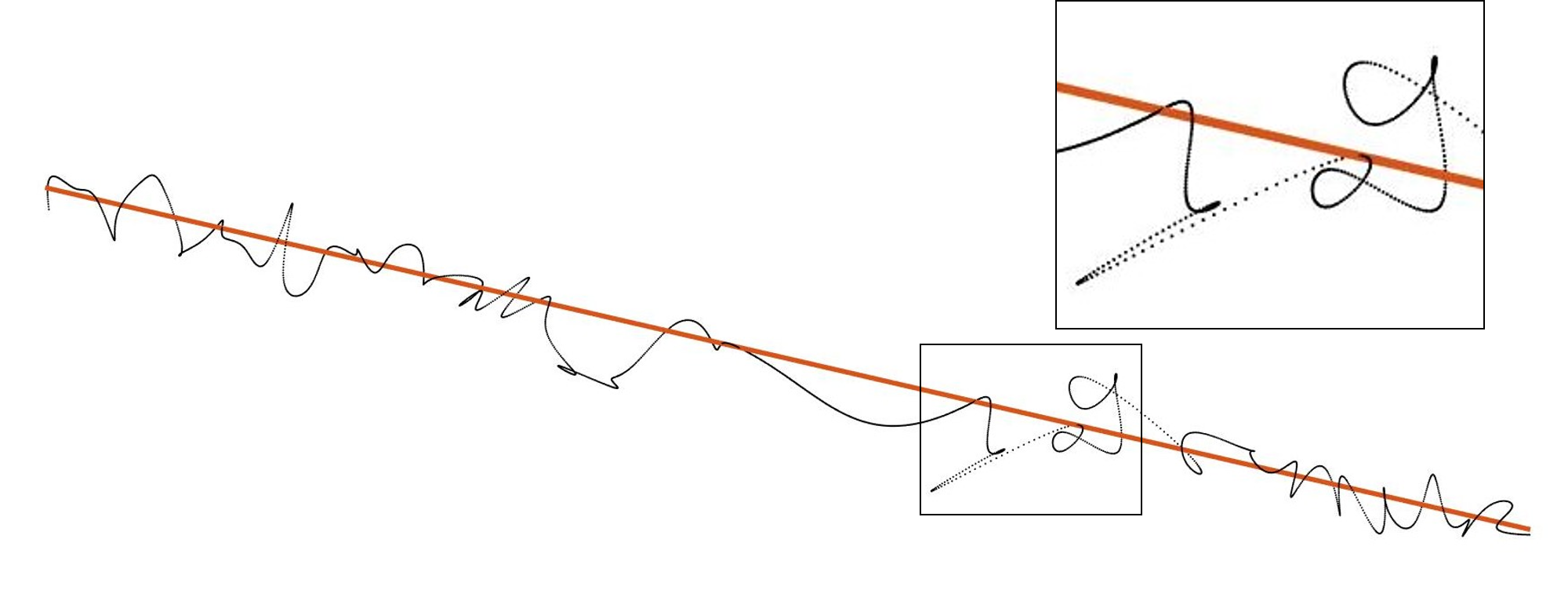}
\caption{A single radial trajectory is shown with its initialization. Notice the density at the high curvature parts of the black curve.}
\label{fig:singleshot}
\end{figure}

\begin{figure}[!ht]
\vspace{-0.2cm}

\centering
\includegraphics[width=0.9\textwidth]{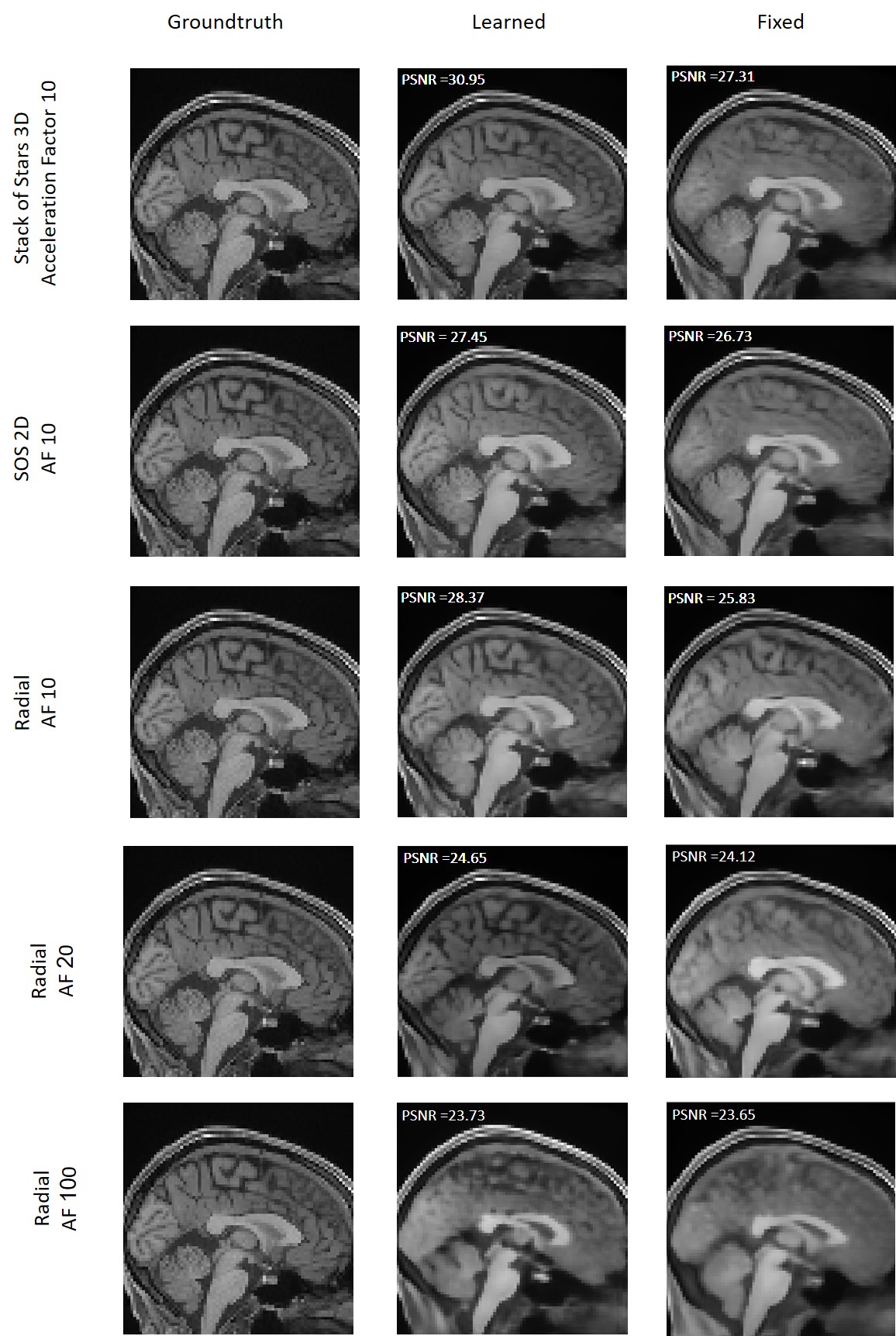}
\caption{Depicted are reconstruction results of all trajectories over different acceleration factors. The images depict a sagittal plane of a sample volume. PSNR is calculated w.r.to the groundtruth image on the left most column.}
\label{saggitalpics}
\end{figure}

\begin{figure}[!ht]
\vspace{-0.2cm}
\centering
\includegraphics[width=0.9\textwidth]{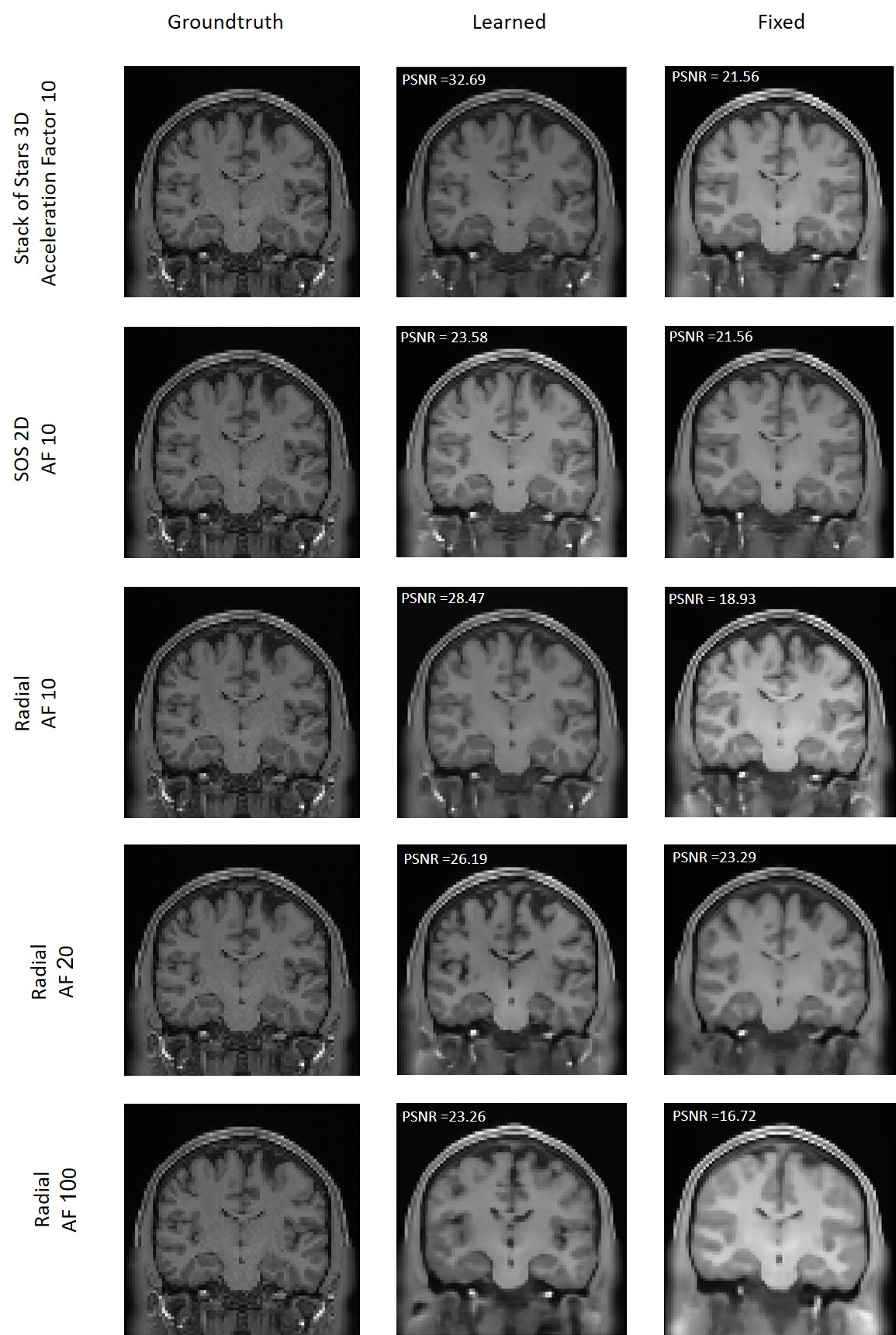}
\caption{Depicted are reconstruction results of all trajectories over different acceleration factors. The images depict a coronal plane of a sample volume. PSNR is calculated w.r.to the groundtruth image on the left most column.}
\label{coronalpics}
\end{figure}

\end{document}